\documentclass[journal=apchd5,manuscript=article]{achemso}

\usepackage[version=3]{mhchem} % Formula subscripts using \ce{}
\usepackage[T1]{fontenc}       % Use modern font encodings
\usepackage{graphicx}		   % Include figure files 
\usepackage{color}
\usepackage{amssymb}
\usepackage{xcolor, soul}
\sethlcolor{yellow}
\author{Volodymyr I. Fesenko}
\affiliation[jilin]
{International Center of Future Science, State Key Laboratory on Integrated Optoelectronics, College of Electronic Science and Engineering, Jilin University, 2699 Qianjin Street, Changchun 130012, China}
\alsoaffiliation[rian]
{Institute of Radio Astronomy of
National Academy of Sciences of Ukraine, 4, Mystetstv Street, Kharkiv 61002, Ukraine}
\email{volodymyr.i.fesenko@gmail.com}
\author{Vitalii I. Shcherbinin}
\alsoaffiliation[jilin]
{International Center of Future Science, State Key Laboratory on Integrated Optoelectronics, College of Electronic Science and Engineering, Jilin University, 2699 Qianjin Street, Changchun 130012, China}
\affiliation[nsc]
{National Science Center `Kharkiv Institute of Physics and Technology' of National Academy of Sciences of Ukraine, 1, Akademicheskaya Str., Kharkiv 61108, Ukraine}
\email{vshch@ukr.net}
\author{Vladimir R. Tuz}
\affiliation[jilin]
{International Center of Future Science, State Key Laboratory on Integrated Optoelectronics, College of Electronic Science and Engineering, Jilin University, 2699 Qianjin Street, Changchun 130012, China}
\alsoaffiliation[rian]
{Institute of Radio Astronomy of
National Academy of Sciences of Ukraine, 4, Mystetstv Street, Kharkiv 61002, Ukraine}
\email{tvr@jlu.edu.cn; tvr@rian.kharkov.ua}

\title[Multiple invisibility regions]
  {Multiple invisibility regions induced by symmetry breaking in a trimer of subwavelength graphene-coated nanowires}


\begin{document}

\newcommand{\onlinecite}[1]{\hspace{-1 ex} \nocite{#1}\citenum{#1}}

%\begin{tocentry}
%\end{tocentry}

\begin{abstract}
Electromagnetic response is studied for clusters of subwavelength graphene-coated nanowires illuminated by a linearly polarized plane wave in the terahertz frequency range. The solution of the scattering problem is obtained with the Lorenz-Mie theory and the multiple cylinder scattering formalism. The results show that normalized scattering cross-sections of nanowire clusters can be drastically changed by the symmetry breaking introduced into the cluster's design. This effect is due to excitation of dark modes and is observed only for the incident wave of TE$_z$-polarization.    
\end{abstract}

\newpage

Electromagnetic scattering of a plane wave by a single or multiple dielectric cylinders is one of the classical problems in the electromagnetic theory and optics \cite{Twersky_JApplPhys_1952, Richmond_AntennaPropag_1965, Olaofe_RadioSci_1970, Magnusson_JOSAA_1989, Lee_JAplPhys_1990, Lee_JQSRT_1992, Bever_ApplOpt_1992}. There is a continuous progress in the development of techniques for solving scattering problems related to various engineering applications. From the practical standpoint, often it is desirable to reduce or cancel waves scattering from the cylindrical object and thus achieve partial or complete `invisibility' effect. This can be realized with special resistive coatings of the cylinders \cite{Sureau_AntennaPropag_1967, Kildal_AntennaPropag_1996, Kim_AntennaPropag_1989}.

Modern progress in nanofabrication techniques and advent of new optical materials have renewed the interest to this research field \cite{Alu_PhysRevLett_2008, Alu_PhysRevB_2009, Jablan_PhysRevB_2009, Chen_ACSNano_2011, Sakhnenko_JSTQE_2013, Bernety_JPCM_2015, Kivshar_Nanoscale_2015,Qian_ACSPhotonics_2018}. For example, with coating made from noble metal at subwavelength scale it becomes possible to reduce scattering from otherwise reflective structures in the infrared and visible parts of spectrum due the excitation of plasmons \cite{Alu_PhysRevLett_2008, Padooru_JApplPhys_2012}. For nanowires made of non-plasmonic materials the conducting coating gives rise to surface plasmons, which are absent in the bare wires \cite{Riso_JOpt_2015}. In visible range, in order to achieve better confinement of surface plasmons, the nanowires typically are coated by noble metals, whereas for the waves in terahertz range it is a common knowledge that graphene is the most suitable coating material \cite{Jablan_PhysRevB_2009}. In the terahertz frequency range graphene shows good conductivity and provides efficient subwavelength confinement \cite{Pengchao_NP_2018}. Remarkably, the conductivity of graphene is high enough to suppress the wave scattering from a coated dielectric nanowire. The tunability of graphene conductivity can be used for an additional control over the location and magnitude of the plasmonic resonances and makes it possible to achieve the invisibility effect for different wavelengths. Notable reduction of the scattering cross-section spectra in the terahertz range has been already demonstrated for both single graphene-coated nanowires \cite{Chen_ACSNano_2011, Riso_JOpt_2015} and clusters of two such nanowires (dimers) \cite{Bernety_JPCM_2015, Naserpour_SR_2017}.

It is known that in clusters of plasmonic nanowires an electromagnetic coupling between constituents becomes possible and can modify essentially the characteristics of wave scattering. The coupling depends on the distance between the nanowires $d$ and the wavelength $\lambda$ of the incident radiation. Four different coupling mechanisms can be distinguished \cite{Savage_Nature_2012, Wei_Chap3_2014}: (i) the far-field coupling takes place between sparsely distributed nanowires, when the distance between the nanowires is larger than the wavelength of the irradiating wave ($d > \lambda$); (ii) the near-field coupling appears when the distance between nanowires (with radius $a$) becomes smaller than the wavelength of the irradiating wave ($d < \lambda$) and lies in the range $\Delta < d \leq 5a$ \cite{Ghosh_ChemRev_2007}, where the critical distance $\Delta~\approx~0.31$~nm is the minimum gap between the nanowires which allows neglecting quantum effects \cite{Savage_Nature_2012}; (iii) the quantum regime occurs for very small distances ($d < \Delta$), where a coherent quantum tunneling of particles becomes possible and the system transits to the state of extreme non-locality in which classical treatments are no longer valid \cite{Savage_Nature_2012, Scholl_NanoLett_2013}; (iv) in the case of contact between nanowires, the lowest-frequency plasmonic mode appears as a true dipole resonance.   

Our interest is in the near-field coupling used to describe hybrid plasmon modes of the cluster. In accordance with this description, the hybrid plasmon modes are treated as an arrangement of the electric dipole moments (or surface charge distributions) induced by the incident radiation for each particle of  the cluster \cite{Hopkins_PhysRevA_2013, Hopkins_PhysRevB_2015}. In such a treatment an analogy with hybridization process of the molecular-orbital theory is often used \cite{Prodan_Science_2003, Chuntonov_NanoLett_2011}, where the hybrid modes are classified in accordance with the point symmetry groups. In particular, in the simplest case of a cluster composed of a chain of two nanowires (dimer), the electric dipole moments of two interacting nanowires can be oriented either co-directionally or counter-directionally and thus can form a symmetrically coupled (bonding) mode or an antisymmetrically coupled (antibonding) mode \cite{Prodan_Science_2003}, respectively. When the electric field of incident wave is polarized along the chain, the plasmons are coupled to form the bonding mode. This leads to a significant red-shift of the plasmon resonance. In the opposite case, when the incident radiation is polarized perpendicular to the chain, the near-fields are in the form of antibonding mode and result in slight blue-shift of the plasmon band. 

It is worth noting that different forms of coupling can appear for more complex clusters having increased number of nanowires. This makes their scattering cross-section much more complicated \cite{Romero_OE_2006, Ghosh_ChemRev_2007, Savage_Nature_2012, Scholl_NanoLett_2013, Wei_Chap3_2014}.
For instance, in the clusters composed of several nanowires the so-called dark modes (also known as trapped modes) can appear. The dark modes are usually referred to any mode that cannot couple directly with an incident wave \cite{Hopkins_PhysRevA_2013, Hopkins_PhysRevB_2015}. Nevertheless, being optically inactive in spatially symmetric clusters, they arise in the clusters with broken symmetry via coupling to an optically active modes (bright modes). Importantly, when such a dark mode is excited, the cluster becomes transparent to the incoming electromagnetic radiation.

In this paper our aim is to demonstrate that several invisibility regions can be realized in a cluster of graphene-coated nanowires due to excitation of dark modes arising from the symmetry breaking in the cluster's design. Trimer is considered as such a cluster, since simpler structures \cite{Riso_JOpt_2015, Chen_ACSNano_2011, Naserpour_SR_2017} lack this peculiarity. 

%-------------------------------------------------%
\section{Problem statement. Scattering formalism}
%--------------------------------------------------
\begin{figure}[tbp]
\centering
\includegraphics[width=0.6\linewidth]{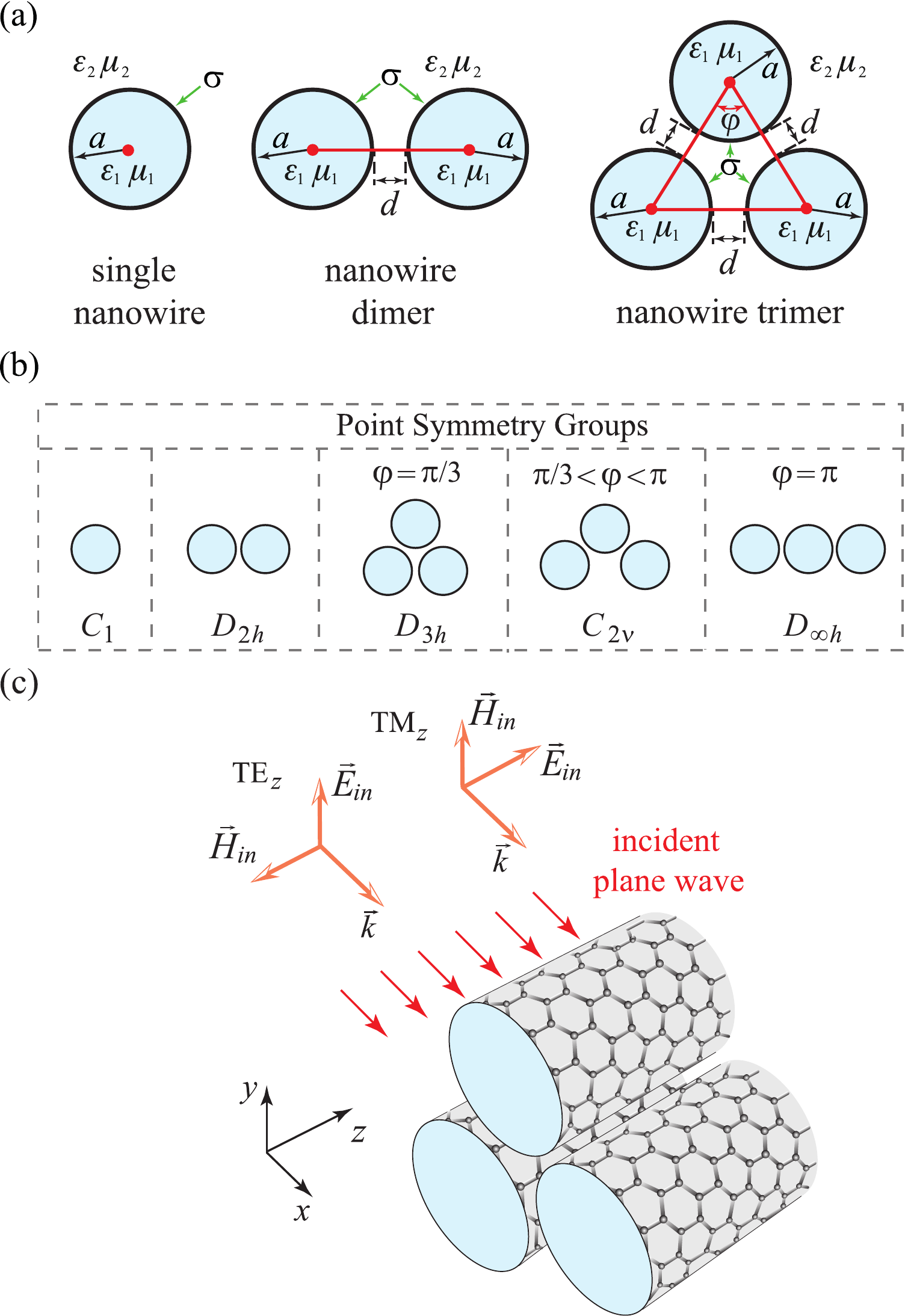}
\caption{(a) Schematic of cross-sections in the $x-y$ plane of a single nanowire as well as nanowire dimer and trimer forming a cluster, (b) classification of the cluster's designs in accordance with the point symmetry groups made in the Sch\"onflies notation, and (c) geometry of the scattering problem where both TE$_z$ and TM$_z$ polarizations of the irradiating (primary) wave are under consideration.}
\label{fig:fig1}
\end{figure}
%--------------------------------------------------

In what follows we compare scattering characteristics of a single nanowire with those of several identical nanowires arranged closely in a cluster (Figure~\ref{fig:fig1}a). Specifically, we consider clusters containing either two nanowires (dimer) or three nanowires (trimer). In the case of a dimer, the nanowires are arranged parallel either to the $x$-axis or  the $y$-axis with the distance $d$ of each other; in a trimer the centers of the nanowires lie at the vertices of a triangle whose vertex angle is $\varphi$ and the base may be oriented either along the $x$-axis or along the $y$-axis.  

All the cluster's designs of interest can be classified according to the point symmetry groups (we use the Sch\"onflies notation \cite{landau_1965_3}; see Figure~\ref{fig:fig1}b): (i) a single nanowire belongs to the trivial symmetry group $C_1$, (ii) a dimer belongs to the dihedral symmetry group $D_{2h}$, and (iii) a symmetric trimer (equilateral triangular cluster; $\varphi=\pi/3$) belongs to the dihedral symmetry group $D_{3h}$, whereas an asymmetric trimer belongs either to the group $C_{2\nu}$ ($\Lambda$-shaped cluster; $\pi/3 < \varphi < \pi$) or to the group $D_{\infty h}$ (linear chain of three nanowires; $\varphi=\pi$).

All nanowires are made from a nonmagnetic ($\mu_1=1$) semiconductor material with the permittivity $\varepsilon_1$. In order to make their scattering characteristics tunable all nanowires are coated by a monolayer graphene sheet having the macroscopic surface conductivity $\sigma$. The nanowires are considered to be infinite along the $z$-axis and have a circular transverse cross-sections in the $x-y$ plane. The  radius of the nanowires is $a$. The subwavelength condition is assumed to be fulfilled, that is, the radii of the nanowires are small enough compared to the wavelength of the incident radiation. The distance $d$ between the nanowires of each cluster is chosen to be less then $2a$, therefore, the resultant condition $d < 2a \ll \lambda$ is satisfied. Finally, the nanowires are embedded into a lossless host medium with the constitutive parameters $\varepsilon_2$ and $\mu_2$.

The wave vector $\vec k$ of the incident electromagnetic wave is directed along the $x$-axis, that is, the nanowires are irradiated orthogonally with respect to their axes. The problem is then reduced to two-dimensional ones, each treating separately for orthogonally polarized waves (Figure~\ref{fig:fig1}c). Thus, we consider the irradiating (primary) wave, which is either TE$_z$-polarized wave with the vectors $\vec E_{in}^{TE}=\{0,E_y,0\}$ and $\vec H_{in}^{TE}=\{0,0,H_z\}$, or TM$_z$-polarized wave with the vectors $\vec E_{in}^{TM}=\{0,0,E_z\}$ and $\vec H_{in}^{TM}=\{0,H_y,0\}$.

The solution of the given scattering problem can be found with the use of the Lorenz-Mie theory and multiple cylinder scattering formalism developed earlier \cite{Twersky_JApplPhys_1952, Lee_JAplPhys_1990, Lee_JQSRT_1992, Schafer_JQSRT_2012}. In this approach, scalar wave potentials are expressed as superpositions of the infinite set of modes. The total wave potentials are derived in the vicinity of each cylinder by summing over all contributions from both the primary wave and the waves scattered by the cylinders. After that, the expressions for the unknown coefficients of the scattered wave potentials are obtained by imposing the required boundary conditions across the surface of each cylinder. They are as follows: (i) tangential components of the total electric field are continuous, and (ii) the discontinuity of the tangential components of the total magnetic field is related to the tangential component of the total electric field via the graphene surface conductivity $\sigma$. 

The resulting system of homogeneous equations for the unknown coefficients of the scattered wave potentials has the same form as that derived for bare (uncoated) nanowires \cite{Lee_JAplPhys_1990}:
%--------------------------------------------------
\begin{equation}
\begin{split}
\sum_{k=1}^{N}~\sum_{s=-\infty}^{+\infty} \left( \delta_{kj} \delta_{ns} + (1-\delta_{kj}) G_{ks}^{jn} \genfrac\{\}{0pt}{0}{b^0_{jnI}}{a^0_{jnII}} \right) \genfrac\{\}{0pt}{0}{b_{ksI}}{a_{ksII}} = \epsilon_j \genfrac\{\}{0pt}{0}{b^0_{jnI}}{a^0_{jnII}}, 
\end{split}
\label{eq:ScatCoeffMult} 
\end{equation}
%--------------------------------------------------
where the conductivity $\sigma$ of graphene coating is introduced in the Mie scattering coefficients $a^0_{jnII}$, $b^0_{jnI}$ and $a_{jnII}$, $b_{jnI}$ for a single graphene-coated nanowire and a cluster of $N$ such nanowires, respectively, $\delta_{kj}$ and $\delta_{ns}$ are the Kronecker delta functions, $\epsilon_j$ is the phase shift relative to the origin of coordinate system for primary wave at the $j$-th nanowire, $ G_{ks}^{jn} = (-i)^{-n+s}H^{(2)}_{-n+s}(k_0R_{kj})\exp{[i(-n+s)\gamma_{kj}]}$, $H^{(2)}_{n}(\cdot)$ is the Hankel function of the second kind, $k_0=\omega/c$ is the wave number in free space, $R_{kj}$ is the distance between centers of $k$-th and $j$-th nanowires, $\gamma_{kj}$ denotes the angular position of the $k$-th nanowire relative to the $j$-th nanowire (for clarity, see Figures~1 and 2 in Ref.~\onlinecite{Lee_JAplPhys_1990}). The explicit expressions of the Mie scattering coefficients for the graphene-coated nanowire are given in Appendix~A.

The scattering cross-sections (SCS) of the cluster can be obtained by averaging fields over all angles of the scattered radiation \cite{Lee_JAplPhys_1990} 
%--------------------------------------------------
\begin{equation}
 C_{sca}^{ij}=\frac{2}{\pi k_0} \int\limits_0^{2\pi} \mid T_{ij}(\theta)\mid^2d\theta,~~~~~~i,j=1,2,
\label{eq:SCSMulti} 
\end{equation}
%--------------------------------------------------
where the elements $T_{ii}$ and $T_{ij}$ of the amplitude scattering matrix correspond to the co-polarized and cross-polarized components of the scattered radiation, respectively. Their expressions are omitted here and can be found in Refs.~\onlinecite{Lee_JAplPhys_1990} and \onlinecite{Lee_JQSRT_1992}. 

Under the normal incidence of the primary wave, the matrix elements responsible for the cross-polarized components of the scattered radiation in \eqref{eq:SCSMulti} reduce to zero. Thus we have:
%--------------------------------------------------
\begin{equation}
\begin{split}
& C_{sca}^\mathrm{TM}=\frac{2}{\pi k_0} \int\limits_0^{2\pi} \mid T_{11}(\theta)\mid^2d\theta, \\
& C_{sca}^\mathrm{TE}=\frac{2}{\pi k_0} \int\limits_0^{2\pi} \mid T_{22}(\theta)\mid^2d\theta.
\label{eq:SCSMultiTETM} 
\end{split}
\end{equation}
%--------------------------------------------------

From \eqref{eq:SCSMultiTETM} the invisibility conditions for the cluster of graphene-coated nanowires can be found. For TM$_z$-polarized and TE$_z$-polarized waves these conditions correspond to the states where $T_{11}(\theta)\to 0$ and $T_{22}(\theta)\to 0$, respectively.

%-----------------------------------------------------------%
\section{Simulation results. Scattering cross-sections and invisibility regions}

To be consistent with results of Refs.~\onlinecite{Riso_JOpt_2015} and \onlinecite{Naserpour_SR_2017}, we perform our study in the wavelength range between $10$ and $60~\mu$m. At these wavelengths the scattering characteristics of bare dielectric nanowire have no peculiarities, so further we consider only nanowires coated by graphene. To give comprehensive analysis, we first compare the scattering characteristics and invisibility regions of a symmetric trimer (design $D_{3h}$) with the available results of Refs.~\onlinecite{Riso_JOpt_2015} and \onlinecite{Naserpour_SR_2017} for a single nanowire (design $C_1$) and a dimer (design $D_{2h}$). Then for the trimer the effect of a symmetry breaking, which is associated with the transition from the trimer design $D_{3h}$ to designs $C_{2v}$ and $D_{\infty h}$, is studied in detail.

In the chosen spectral range the complex conductivity of graphene is well described by the Kubo formula (see Appendix~B). The nanowires are considered to be made from SiO$_2$. In accordance with tabular data of Ref.~\onlinecite{Palik_book_1998}, in the wavelength range under consideration (which corresponds to the photon energies $0.02-0.12$~eV), the material losses in SiO$_2$ are small enough and thus can be ignored. Therefore, in our numerical model we account for the Ohmic losses in graphene, but neglect material losses in the nanowires as well as spatial dispersion of graphene.

%--------------------------%
\subsection{Symmetric designs. Formation of an additional invisibility region}

\begin{figure}[ht!]
\centering
\includegraphics[width=0.6\linewidth]{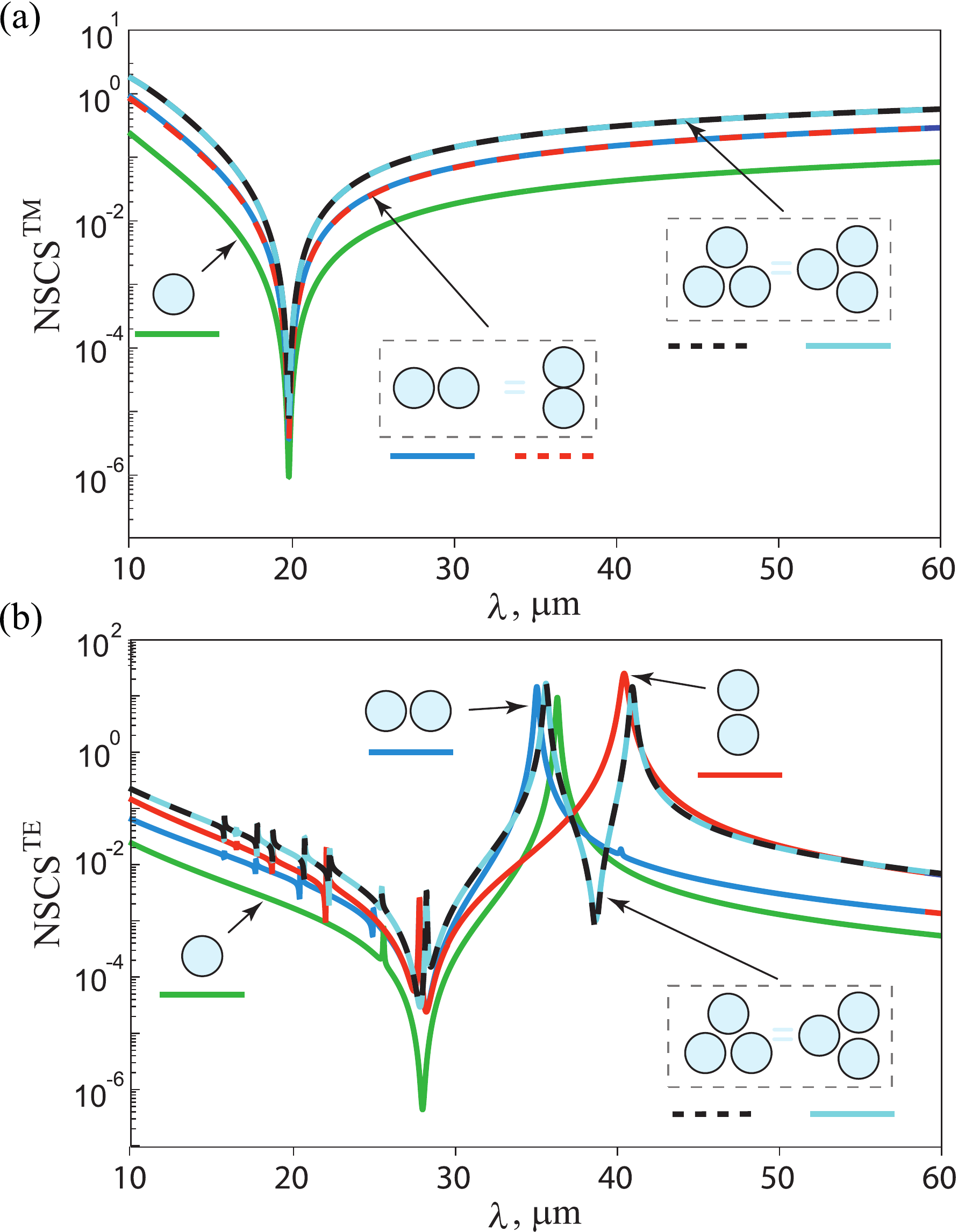}
\caption{Spectral curves of (a) NSCS$^\mathrm{TM}$ and (b) NSCS$^\mathrm{TE}$ of a single graphene-coated nanowire ($C_1$), a cluster of two graphene-coated nanowires (dimer, $D_{2h}$), and a cluster of three graphene-coated nanowires (symmetric trimer, $D_{3h}$). Different orientation of clusters with respect to the propagation direction of incident wave are considered. Here, we set $a=0.5~\mu$m, $\varepsilon_1=3.9$~(SiO$_2$), $\varepsilon_2=1.0$~(air), $\mu_1=\mu_2=1.0$, $T=300$~K, $\mu_c=0.5$~eV, $\Gamma=0.1$~meV. The distance between individual nanowires in clusters is $d=50$~nm.}
\label{fig:fig2}
\end{figure}

For most plasmonic structures the scattering spectra depend strongly on the polarization of the incident electromagnetic radiation \cite{Ghosh_ChemRev_2007, Wei_Chap3_2014}. This is also true for graphene-coated nanowires and can be seen from Figure~\ref{fig:fig2}, which shows the spectral features of the normalized scattering cross-section (NSCS; normalized on the nanowire's diameter $2a$)  for structures illuminated by both TM$_z$-polarized and TE$_z$-polarized primary waves. 

For a $C_1$-nanowire, $D_{2h}$-dimer, and $D_{3h}$-trimer illuminated by TM$_z$-polarized wave, the curves of NSCS (NSCS$^\mathrm{TM}$) behave similarly and all exhibit a single invisibility region near $\lambda\approx 19.9~\mu$m (Figure~\ref{fig:fig2}a). Such spectral behavior is typical for the graphene coated nanowire \cite{Arruda_JOSAB_2014} and is due to the excitation of a bulk plasmon resonance in the nanowire core (see also Figure S2 in Supporting Information). The small difference between the curves consists in slight increase of the scattering level and narrowing of the invisibility region with increasing number $N$ of nanowires in the cluster. In the case of TM$_z$-polarized incident wave similar scattering characteristics were reported for a cluster of elliptical objects covered by patterned graphene \cite{Bernety_JPCM_2015}. 

%--------------------------------------------------
%--------------------------------------------------

For the $C_1$-nanowire and both cluster's designs the local minima of the NSCS$^\mathrm{TM}$ correspond to the wavelengths, where the numerator of the coefficient $b^0_{1I}$ in Equation~(\ref{eq:ScatEffNorm}) and the element $T_{11}$ of amplitude scattering matrix in Equation~(\ref{eq:SCSMultiTETM}) reduce to zero, respectively. As mentioned above, these wavelengths are nearly the same and can be estimated as $\lambda_{inv}^\mathrm{TM} = \lambda_{inv}/\sqrt{2}$ \cite{Naserpour_SR_2017}, where $\lambda_{inv} = 2\pi c \left[ e^2\mu_c/(\pi\hslash^2\varepsilon_0 a(\varepsilon_1-\varepsilon_2))\right]^{-\frac{1}{2}}$.
% %--------------------------------------------------
% \begin{equation}
% \lambda_{inv} = 2\pi c \left[ \frac{e^2\mu_c}{\pi\hslash^2\varepsilon_0 a(\varepsilon_1-\varepsilon_2)}\right]^{-\frac{1}{2}}.
% \label{eq:LamInv1}
% \end{equation}
% %--------------------------------------------------

For TE$_z$-polarized incident wave the resultant NSCS spectra (NSCS$^\mathrm{TE}$) have more complicated structure  (Figure~\ref{fig:fig2}b), especially for clusters, where the coupling between individual nanowires becomes evident. For a single $C_1$-nanowire the NSCS$^\mathrm{TE}$ includes the invisibility region at the wavelength $\lambda_{inv}\approx28~\mu$m, where the coefficient $a^0_{1II}$ in Equation~(\ref{eq:ScatEffNorm}) is zero. Besides, at shorter wavelengths there are several plasmonic resonant peaks, which are related to the complex poles of the high-order (multi-pole) coefficients $a^0_{nII}$ of the electromagnetic field expansion (see, for instance, the near field pattern shown in Figure~S2 of Supporting Information for $\lambda=25.7~\mu$m). The main resonant peak corresponds the pole of $a^0_{1II}$, and lies close to wavelength \cite{Naserpour_SR_2017} $\lambda_{res} = 2\pi c \left[ e^2\mu_c/(\pi\hslash^2\varepsilon_0 a(\varepsilon_1+\varepsilon_2))\right]^{-\frac{1}{2}}$.
% %--------------------------------------------------
% \begin{equation}
% \lambda_{res} =  2\pi c \left[\frac{e^2\mu_c}{\pi\hslash^2\varepsilon_0 a(\varepsilon_1+\varepsilon_2)}\right]^{-\frac{1}{2}}.
% \label{eq:LamInv2}
% \end{equation}
% %--------------------------------------------------
For given parameters of the $C_1$-nanowire coated by graphene $\lambda_{res}\approx36~\mu$m (see Figure~\ref{fig:fig2}b).

The $D_{2h}$-dimer features another position of the main resonant peak. As was found in Refs.~\onlinecite{Naserpour_SR_2017}, \onlinecite{Wei_Chap3_2014}, and \onlinecite{Romero_OE_2006}, the NSCS$^\mathrm{TE}$ spectrum of dimer depends on its spatial orientation with respect to the direction of wave incidence. Our calculations (Figure~\ref{fig:fig2}b) confirm this peculiarity. We consider two principal orientations of the $D_{2h}$-dimer, where the chain of wires is directed either along the $x$-axis or along the $y$-axis. Note that the axes of the individual nanowires in cluster are always oriented parallel to the $z$-axis. From Figure~\ref{fig:fig2}b follows that for both spatial orientations of dimer the spectral position of the invisibility region is close to that ($\lambda\approx28~\mu$m) for a single nanowire. For dimers, the strong coupling between Mie-modes of individual nanowires gives rise to additional secondary resonance inside the invisibility region. Such resonance is unwanted because it partially suppresses the invisibility effect. Fortunately it can be spectrally shifted far away from the invisibility region by increasing the distance $d$ between nanowires in the cluster.

%--------------------------------------------------
\begin{figure}[ht!]
\centering
\includegraphics[width=0.6\linewidth]{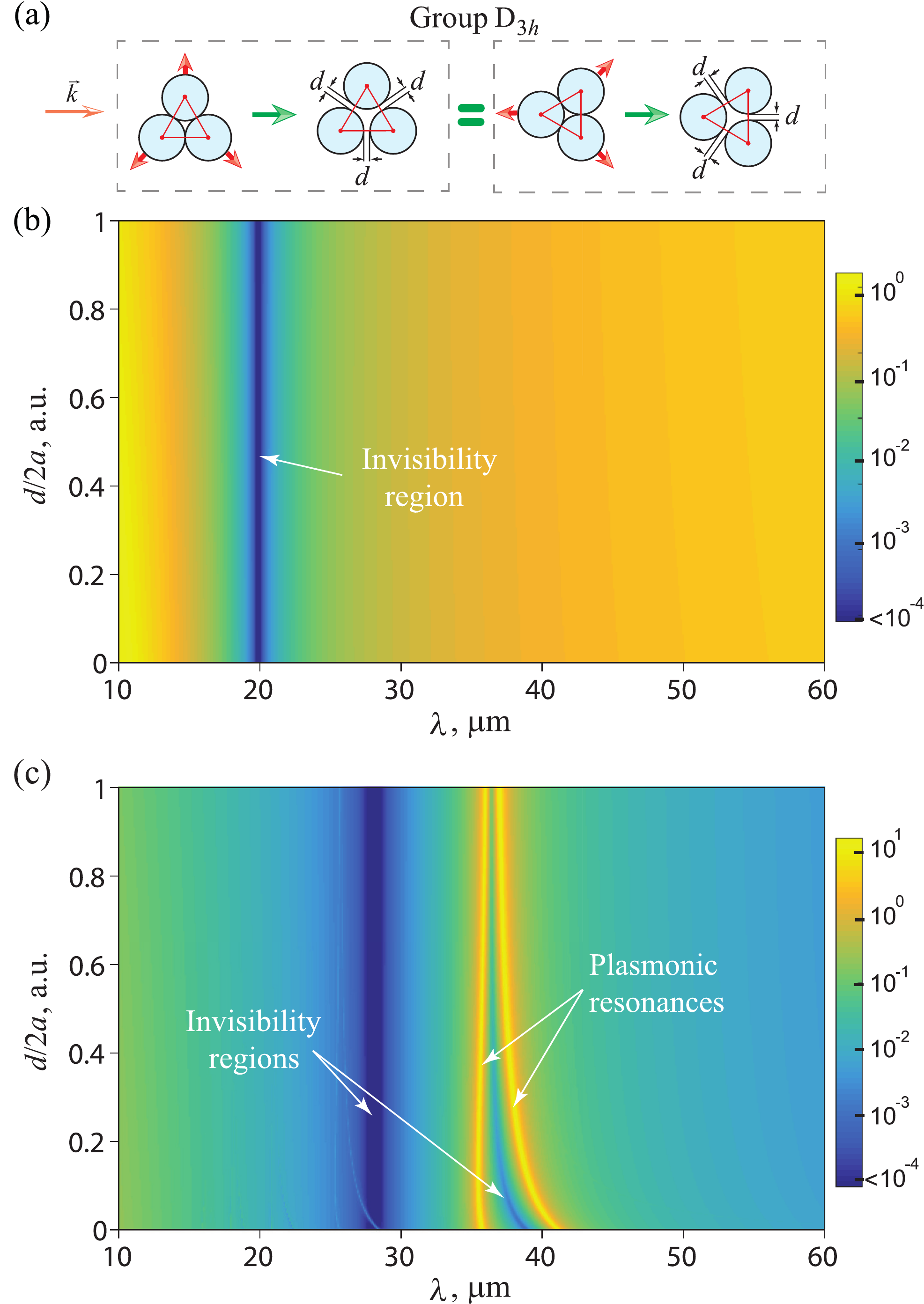}
\caption{(a) Schematic of two arrangements of three graphene-coated nanowires (symmetric trimer, $D_{3h}$) with respect to the propagation direction of irradiating wave. The spectra of normalized scattering cross-section versus the distance $d$ between nanowires for the $D_{3h}$-trimer irradiated by (b) TM$_z$-polarized wave, and (c) TE$_z$-polarized wave. All parameters of the problem are the same as in Figure~\ref{fig:fig2}.}
\label{fig:fig3}
\end{figure}
%--------------------------------------------------
In contrast to behavior of the invisibility region, the spectral positions of the main and secondary plasmonic resonances are strongly depended on the spatial orientation of the $D_{2h}$-dimer. In the first case, the NSCS$^\mathrm{TE}$ spectrum (red solid line in Figure~\ref{fig:fig2}b) demonstrates the strong red shift in comparison with that of the $C_1$-nanowire (green solid line). Contrary to this, for the second orientation of the $D_{2h}$-dimer the slight blue-shift is observed for the spectral position of the resonant peak (blue solid line). Besides, in the latter case, all of the secondary plasmonic resonances are blue-shifted relative to those in the former case. Similar spectral characteristics for dimers of metal-coated and graphene-coated dielectric nanocylinders were reported in Refs. \onlinecite{Naserpour_SR_2017} and \onlinecite{Romero_OE_2006}, respectively. Note that the scattering characteristics shown in Figure~\ref{fig:fig2} for the dimers agree closely with those calculated in Ref.~\onlinecite{Naserpour_SR_2017} by using COMSOL Multiphysics software package. This lends support to the validity of our analytical treatment for the multiple nanowires scattering.   

Aggregation of graphene-coated nanowires into a trimer gives rise to new forms of mode interactions between the cluster's constituents and thereby drastically changes the overall scattering characteristics even for the symmetric $D_{3h}$-trimer. In fact, plasmons in the trimer appear in hybridized states and produce resonant response at particular wavelengths. For given cluster consisting of subwavelength elements, the dipole approximation can be used to describe the hybrid plasmonic states. In this approximation, for each nanowire only two dipole terms of the expansion (\ref{eq:ScatCoeffMult}) can be considered \cite{Quinten_SurfSci1986, Hopkins_PhysRevB_2015}, i.e. nonzero Mie scattering coefficients are $a_1$ (electric dipole) and $b_1$ (magnetic dipole) (for trimer possible orientations of electric dipole moments are shown in Table S1 in Supporting Information by red arrows). 

Among trimers, the symmetric one in the form of equilateral triangle features the highest symmetry $D_{3h}$ (Figure~\ref{fig:fig1}b). Owing to this, the NSCS spectra of $D_{3h}$-trimer are independent of its spatial orientation with respect to the direction of wave incidence for both  TM$_z$-polarized and TE$_z$-polarized waves. This can be seen from Figures~\ref{fig:fig2}a,b and \ref{fig:fig3}b,c. As was discussed above, the NSCS$^\mathrm{TM}$ spectrum of the $D_{3h}$-trimer has a simple form and remains invariant with the distance $d$ between the individual nanowires (Figure~\ref{fig:fig3}b). This derives from the fact that this spectrum does not involve plasmonic resonances. Thus plasmonic coupling between several nanowires occurs for the TE$_z$-polarized incident wave only. For this reason, in the following we restrict consideration to the case of TE$_z$-polarization.  

Figure~\ref{fig:fig2}b shows the NSCS$^\mathrm{TE}$ spectra for the $D_{3h}$-trimer. In addition, Figure~\ref{fig:fig3}c also demonstrates the impact of the distance $d$ between nanowires on the resulting NSCS$^\mathrm{TE}$ spectrum. It can be seen that the main invisibility region discussed above is relatively unaffected by $d$, whereas the major changes are observed in the long wavelength region of plasmon resonances. In this region interaction between plasmon modes initiates splitting of the single plasmon resonance into two ones. The splitting appears due to the hybridization of the dipole moments having specific orientation in accordance with the sub-group $E'$ of the symmetry group $D_{3h}$ (see Table S1 in Supporting Information). Here the first hybrid state corresponds to the shorter wavelength $E'$ antibonding mode, whereas the second one corresponds to the longer wavelength $E'$ bonding mode (to be more specific, when the trimer's base is oriented parallel to the propagation direction of the incident wave, the shorter and  longer wavelength plasmon resonances correspond to the modes presented in rows 3 and 4 of Table S1, respectively; for the orthogonal trimer's orientation these plasmon resonances correspond to the modes shown in rows 2 and 5 of Table S1, respectively). In all cases the hybrid $E'$ modes have nonzero total dipole moments, therefore such modes are optically active (i.e. they are bright modes) and can efficiently couple to the field of the irradiating wave at the corresponding resonant wavelengths. These resonant wavelengths are located near those of plasmon resonances for the $D_{2h}$-dimers of two basic orientations.

We should note that our finding for the hybrid plasmon modes of the $D_{3h}$-trimer of graphene-coated nanowires are in good agreement with those of Ref.~\onlinecite{Chuntonov_NanoLett_2011} for a trimer of plasmon (uncoated) spherical nanoparticles. 

Importantly, in a gap between these two plasmon resonances an additional invisibility region of the NSCS$^\mathrm{TE}$ appears. It arises due to the strong near-field coupling of neighboring graphene-coated nanowires. While the spectral position and bandwidth of the main invisibility region remains unchanged with increasing distance $d$ between individual nanowires, additional invisibility region therewith changes distinctly. As $d$ increases, the bandwidth of the latter region is narrowed, since the shorter wavelength and longer wavelength resonant peaks approach each other and tend to that of uncoupled (individual) nanowires. 
 
%-----------------------------%
\subsection{Asymmetric designs. Formation of multiple invisibility regions}

\begin{figure}[ht!]
\centering
\includegraphics[width=0.6\linewidth]{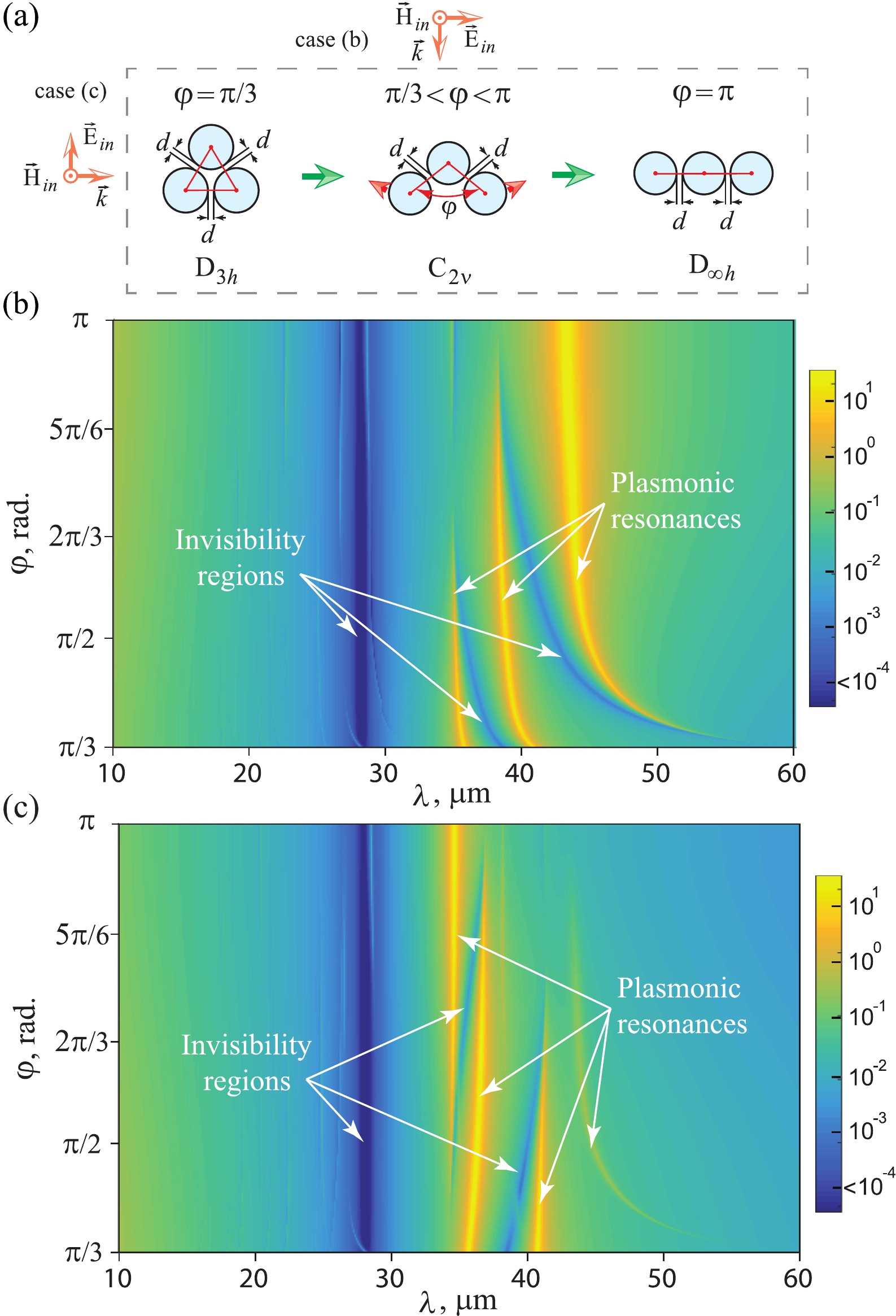}
\caption{(a) Illustration of the symmetry lowering of trimer denoted in the terms of the point symmetry groups ($D_{3h}\to C_{2\nu}\to D_{\infty h}$). It is performed by changing the vertex angle $\varphi$ of the triangle which defines the geometry of cluster. The NSCS$^\mathrm{TE}$ spectra of the asymmetric trimer as function of vertex angle, where the base of triangle is oriented along (b) the $x$-axis and (c) the $y$-axis. Here we set $d=10$~nm. All other parameters of the problem are the same as in Figure~\ref{fig:fig2}. }
\label{fig:fig4}
\end{figure}

More invisibility regions may appear in the case of nanowire trimers with broken symmetry, which disturbs the arrangement of dipole moments in the cluster and provides an additional route for excitation of plasmon hybridized states. By analogy with Ref.~\onlinecite{Chuntonov_NanoLett_2011}, we study the structural symmetry breaking, which arises from the opening the vertex angle $\varphi$ for the triangle forming the trimer design (Figure ~\ref{fig:fig4}a). 

The impact of the symmetry lowering ($D_{3h}\to C_{2\nu}\to D_{\infty h}$) on NSCS spectra is shown in Figures~\ref{fig:fig4}b and \ref{fig:fig4}c for two orthogonal orientations of trimer with respect to the propagation direction of the  TE$_z$-polarized incident wave. The increase of the vertex angle $\varphi$ (i.e. $\varphi>\pi/3$) removes degeneracy of plasmon modes. As a result, new plasmon resonances and invisibility regions arise in the NSCS$^\mathrm{TE}$ spectra. For each angle $\varphi$, the number and spectral positions of plasmon resonant peaks and invisibility regions depend on spatial orientation of the trimer. Moreover, for some $\varphi$, trimers of different orientation may have completely different spectral characteristics. When the base of the triangle, which forms the trimer, is oriented perpendicular (Figure~\ref{fig:fig4}b) or parallel (Figure~\ref{fig:fig4}c) to the wave vector $\vec{k}$ of the incident wave, then all resonance bands and invisibility regions experience blue-shift or red-shift with increasing $\varphi$, respectively. 

In contrast to the wavelength ($\lambda\approx 28~\mu$m) of the main invisibility region, which is independent of asymmetry and orientation of the trimer, central wavelengths of additional invisibility regions for the $C_{2\nu}$-trimer can be controlled by changing the vertex angle $\varphi$ as shown in Figures~\ref{fig:fig4}b and \ref{fig:fig4}c. 

Mathematically speaking, when the vertex angle $\varphi$ gradually increases from $\pi/3$ to $\pi$, the sub-group $E'$ of the symmetry group $D_{3h}$ transforms to the sub-groups $A_1$ and $B_2$ of the group $C_{2\nu}$. These $A_1$ and $B_2$ states are distinct for the bonding and antibonding modes, respectively (see Table S1 and Figure S3 in Supporting Information). As soon as an asymmetry is introduced to the trimer design, the dark modes gain the capacity to couple to the field of the incident wave and thus can be excited. Indeed, the dark ringlike state of the sub-group $A'_2$ of the group $D_{3h}$ transforms to the bright state with non-zero dipole moment belonging to the sub-group $B_2$ of the group $C_{2\nu}$.

%--------------------------------------------------
%--------------------------------------------------

For $\varphi=\pi$ the cluster reduces to the final geometry of the group $D_{\infty h}$, which represents a chain of three nanowires. There are only single resonant peak and single invisibility region for both orthogonal orientations of the nanowire cluster. When the chain of the $D_{\infty h}$-trimer is oriented perpendicular to the direction of the incident wave, the $B_2$ bonding mode transforms to the bright $\varSigma_u^+$ bonding mode with non-zero total dipole moment, which is oriented along the $y$-axis. Two another states are dark modes for the $D_{\infty h}$-trimer (Figure~\ref{fig:fig4}b). At the same time, for the $D_{\infty h}$-trimer whose chain is oriented along the direction of  wave incidence only the $\Pi_u$ antibonding mode appears as a result of transformation from the sub-group $A_1$ of the group $C_{2\nu}$ (Figure~\ref{fig:fig4}c). 

In fact, the NSCS spectra of the $D_{\infty h}$-trimer have much in common with those of the the $D_{2h}$-dimer oriented in the same direction and acquire only a small red-shift of the resonant peaks. This effect was firstly reported for the chains of gold spheres \cite{Quinten_SurfSci1986}.

\section{Conclusions}

Within the Lorenz-Mie theory and multiple cylinder scattering formalism, scattering characteristics has been investigated for a cluster of three graphene-coated nanowires (trimer) illuminated by a linearly polarized plane wave in the terahertz frequency range. Both TE$_z$-polarized and TM$_z$-polarized incident waves were considered. Comparison has been performed with known results for an isolated graphene-coated nanowire and a dimer. In the case of TM$_z$-polarized incident wave the NSCS spectra of all above-mentioned structures behave similarly and exhibit a single invisible region. No resonant peaks have been found in the spectra, which is consistent with the fact that surface plasmons of this polarization are nonexistent.   

In the case of TE$_z$-polarized incident wave, the NSCS spectra of the single nanowire ($C_1$) and both $D_{2h}$ (dimer) and $D_{3h}$ (symmetric trimer) nanowire clusters differ noticeably from each other. In clusters the strong coupling between plasmon modes of individual nanowires gives rise to several plasmonic resonances and invisibility regions in NSCS spectra. For instance, there are two invisibility regions and two main resonant peaks in the NSCS spectrum of the $D_{3h}$-trimer. Moreover such scattering spectrum is independent from spatial orientation of the trimer due to high symmetry of the group $D_{3h}$. 

The impact of the symmetry lowering (which is associated with the transition from the design $D_{3h}$ to designs $C_{2v}$ and $D_{\infty h}$) in the graphene-coated nanowire trimer on its NSCS spectrum has been clearly shown. In NSCS spectra of the asymmetric trimers an additional invisibility region has been found to exist. Such region arises due to excitation of a dark mode. 

An additional advantage of graphene-coated nanowires is ability to tune the main and additional invisibility regions in a wide frequency range by modifying the chemical potential of graphene with chemical doping or electrical gating.

\section*{Appendix A. Scattering cross-sections of a~single nanowire}
\label{sec:AppB}
%--------------------------------------------------
\renewcommand{\theequation}{A.\arabic{equation}}
\setcounter{equation}{0}
%--------------------------------------------------

The scattering efficiencies $Q_{sca}^\mathrm{TE}$ and $Q_{sca}^\mathrm{TM}$ of a single graphene-coated nanowire can be analytically derived from the Lorenz-Mie scattering theory \cite{Bohren_book_1998, Chen_ACSNano_2011, Riso_JOpt_2015, Naserpour_SR_2017}. Using standard Mie expansion, the  field scattered by the nanowire is presented as a discrete sum of $n$ cylindrical harmonics having complex amplitudes. In the case of small nanowire radius, we can take into account only the first (dipolar) terms of the Mie expansion \cite{Bohren_book_1998, Alu_PhysRevE_2005}, expressing the resulting scattering efficiencies as follows:
%--------------------------------------------------
\begin{equation}
\begin{split}
& Q_{sca}^\mathrm{TE} = \frac{2}{k_0a}\left[ \mid a^0_{0II}\mid^2 + 2\sum_{n=1}^\infty \left( \mid a^0_{nII}\mid^2 + \mid b^0_{nII}\mid^2 \right) \right],\\
& Q_{sca}^\mathrm{TM} = \frac{2}{k_0a}\left[ \mid b^0_{0I}\mid^2 + 2\sum_{n=1}^\infty \left( \mid b^0_{nI}\mid^2 + \mid a^0_{nI}\mid^2 \right) \right],
\end{split}
\label{eq:ScatEff} 
\end{equation}
for the TE$_z$-polarized and TM$_z$-polarized waves, respectively. 
The Mie scattering coefficients  are \cite{Chen_ACSNano_2011, Riso_JOpt_2015, Naserpour_SR_2017}:
%--------------------------------------------------
\begin{equation}
\begin{split}
&a^0_{nI} = \frac{J_n(\xi)H'^{(1)}_n(\xi) - J'_n(\xi)H^{(1)}_{n}(\xi)}{J_n(\eta)H'^{(1)}_{n}(\xi) - m J'_n(\eta)H^{(1)}_{n}(\xi) + i\sigma Z_0 J_n(\eta)H^{(1)}_{n}(\xi)} ,\\
&b^0_{nI} =  \frac{J_n(\eta)J'_n(\xi) - m J'_n(\eta)J_n(\xi)+i Z_0 \sigma J_n(\eta)J_n(\xi)} {J_n(\eta)H'^{(1)}_{n}(\xi) - m J'_n(\eta)H^{(1)}_{n}(\xi) + i Z_0 \sigma J_n(\eta)H^{(1)}_{n}(\xi)} ,\\
&a^0_{nII} = \frac{m J_n(\eta)J'_n(\xi) - J'_n(\eta) J_n(\xi) + i Z_0 \sigma J'_n(\eta) J'_n(\xi)}{m J_n(\eta)H'^{(1)}_{n}(\xi) - J'_n(\eta)H^{(1)}_{n}(\xi) + i Z_0 \sigma J'_n(\eta) H'^{(1)}_n(\xi)}, \\
&b^0_{nII} =  \frac{m \left[ J_n(\xi)H'^{(1)}_n(\xi) - J'_n(\xi)H^{(1)}_{n}(\xi) \right]}{m J_n(\eta)H'^{(1)}_{n}(\xi) - J'_n(\eta)H^{(1)}_{n}(\xi) + i Z_0 \sigma J'_n(\eta) H'^{(1)}_n(\xi)},
\end{split}
\label{eq:ScatCoeff} 
\end{equation}
%--------------------------------------------------
where $\eta = k_1a$, $\xi = k_2a$, $k_1=k_0\sqrt{\varepsilon_1 \mu_1}$, $k_2=k_0\sqrt{\varepsilon_2 \mu_2}$, $Z_0=\sqrt{\mu_0/\varepsilon_0}$ is the impedance of free space, $m=\sqrt{\varepsilon_1/\varepsilon_2}$ is the relative refractive index of a single nanowire, $J_n(\cdot)$ and $J'_n(\cdot)$ are the Bessel function of the first kind and its derivative with respect to the function argument, and $H^{(1)}_n(\cdot)$ and $H'^{(1)}_n(\cdot)$ are the Hankel function of the first kind and its derivative, respectively.

If the nanowire is under the normal wave incidence, the Mie scattering coefficients $a^0_{nI}$ and $b^0_{nII}$ vanish in Equation~(\ref{eq:ScatEff}) and expressions for the scattering efficiencies can be simplified:  
%--------------------------------------------------
\begin{equation}
\begin{split}
& Q_{sca}^\mathrm{TE} = \frac{2}{k_0a}\sum_{n=-\infty}^{+\infty} \mid a^0_{nII}\mid^2=\frac{1}{2a}C_{sca}^{TE},\\
& Q_{sca}^\mathrm{TM} = \frac{2}{k_0a}\sum_{n=-\infty}^{+\infty} \mid b^0_{nI}\mid^2=\frac{1}{2a}C_{sca}^{TM},
\end{split}
\label{eq:ScatEffNorm} 
\end{equation}
%--------------------------------------------------
here $C_{sca}^\mathrm{TE}$ and $C_{sca}^\mathrm{TM}$ are the scattering cross-sections (SCS) for the TE$_z$-polarized and TM$_z$-polarized waves, respectively. The invisibility conditions are $a^0_{nII}\to 0$ and $b^0_{nI}\to 0$ for the TE$_z$-polarized and TM$_z$-polarized waves, respectively. The frequencies of plasmonic resonances, if any, correspond to the poles of the Mie scattering coefficients.    

\section*{Appendix B. Graphene description}
\label{sec:AppC}
%--------------------------------------------------
\renewcommand{\theequation}{B.\arabic{equation}}
\setcounter{equation}{0}
%--------------------------------------------------
Ignoring the quantum finite-size effect of graphene, the nanowire coating is treated as an infinitely thin graphene sheet having the macroscopic surface conductivity $\sigma$ dependent on the angular frequency $\omega=2\pi f$, chemical potential $\mu_c$, ambient temperature $T$, and charge carries scattering rate $\Gamma$. The surface conductivity of graphene consists of intraband and interband contributions $\sigma = \sigma_{intra} + \sigma_{inter}$, which are described by Kubo formalism \cite{Falkovsky_PhysRev_2007}: 
\begin{equation}
\begin{split}
& \sigma_{intra} = \frac{2 i e^2k_BT}{\hslash^2 \pi \left(\omega+i \Gamma\right)}\ln\left[2\cosh\left(\frac{\mu_c}{2k_BT} \right) \right],\\
&\sigma_{inter} = \frac{e^2}{4\hslash\pi}
\left[\frac{\pi}{2}+\arctan\left(\frac{\hslash\omega-2\mu_c}{2k_BT}\right) -\right.\\
&-\left.\frac{i}{2}\ln\frac{\left(\hslash\omega+2\mu_c\right)^2}{\left(\hslash\omega-2\mu_c\right)^2+\left(2k_BT\right)^2} \right].
\end{split}
\label{eq:Kubo} 
\end{equation}
Here $k_B$ is the Boltzmann constant, $\hslash$ is the reduced Planck constant, and $e$ is the electron charge. The chemical potential $\mu_c$  is related to the carriers density $N_c$ as $\mu_c=\hslash v_F \sqrt{\pi N_c}$, where $v_F\simeq 10^6$~m/s is the Fermi velocity of electrons in graphene. 

%--------------------------------------------------------
\section*{Associated content}
\subsection*{Supporting Information}
Table S1 represents arrows of dipole plasmon modes of trimers obtained from the plasmon hybridization theory; Figures S2 and S3 demonstrate the near field patterns which supplement Figures~\ref{fig:fig2}b, \ref{fig:fig3}c and Figures~\ref{fig:fig4}b, \ref{fig:fig4}c of the manuscript, respectively.

%--------------------------------------------------------
\section*{Author information}
\subsection*{Corresponding Author}

*E-mail: tvr@jlu.edu.cn; tvr@rian.kharkov.ua.

\subsection*{Author Contributions}
All authors contributed equally to this work. All authors have given approval to the final version of the manuscript.

\subsection*{ORCID}
Volodymyr I. Fesenko: 0000-0001-9106-0858\\
Vitalii I. Shcherbinin: 0000-0002-9879-208X\\
Vladimir R. Tuz: 0000-0001-6096-7465

\subsection*{Notes}

The authors declare no competing financial interest.

%--------------------------------------------------------
\begin{acknowledgement}

The authors acknowledge Jilin University for hospitality and financial support. 

\end{acknowledgement}


\bibliography{nanowires_trimer}

\newpage
\section*{For Table of Contents Use Only}

\subsection*{Manuscript title:} Multiple invisibility regions induced by symmetry breaking in a trimer of subwavelength graphene-coated nanowires

\subsection*{Names of authors:} Volodymyr I. Fesenko, Vitalii I. Shcherbinin, Vladimir R. Tuz

\subsection*{Brief synopsis:}
Electromagnetic response is studied for clusters of subwavelength graphene-coated nanowires illuminated by a linearly polarized plane wave in the terahertz frequency range. The solution of the scattering problem is obtained with the Lorenz-Mie theory and the multiple cylinder scattering formalism. The results show that normalized scattering cross-sections of nanowire clusters can be drastically changed by the symmetry breaking introduced into the cluster's design. This effect is due to excitation of dark modes and is observed only for the incident wave of TE$_z$-polarization.

\begin{figure}[ht!]
\centering\includegraphics[width=6cm]{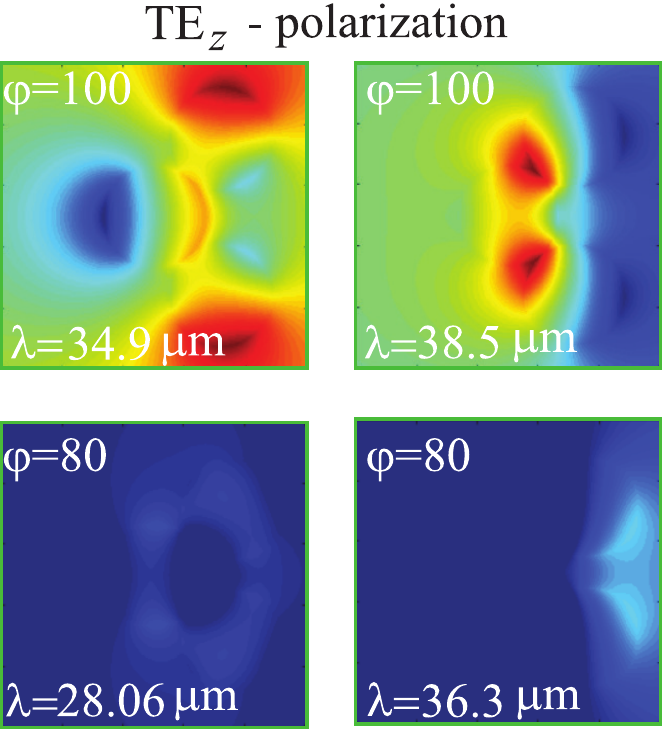}
\caption*{ToC-image}
\label{fig:toc}
\end{figure}

\subsection*{Supporting Information:}

This Supporting Information consists of Table S1 with data for dipolar plasmon modes of trimers with varying vertex angle $\pi/3\le \varphi \le \pi$, followed from the plasmon hybridization theory \cite{Chuntonov_NanoLett_2011}; Figure S2 with the near field patterns supplementing Figures 2b and 3c of the manuscript; Figure S3 with the near field patterns supplementing Figure 4 of the manuscript.

\begin{figure}[ht!]
\centering\includegraphics[width=10cm]{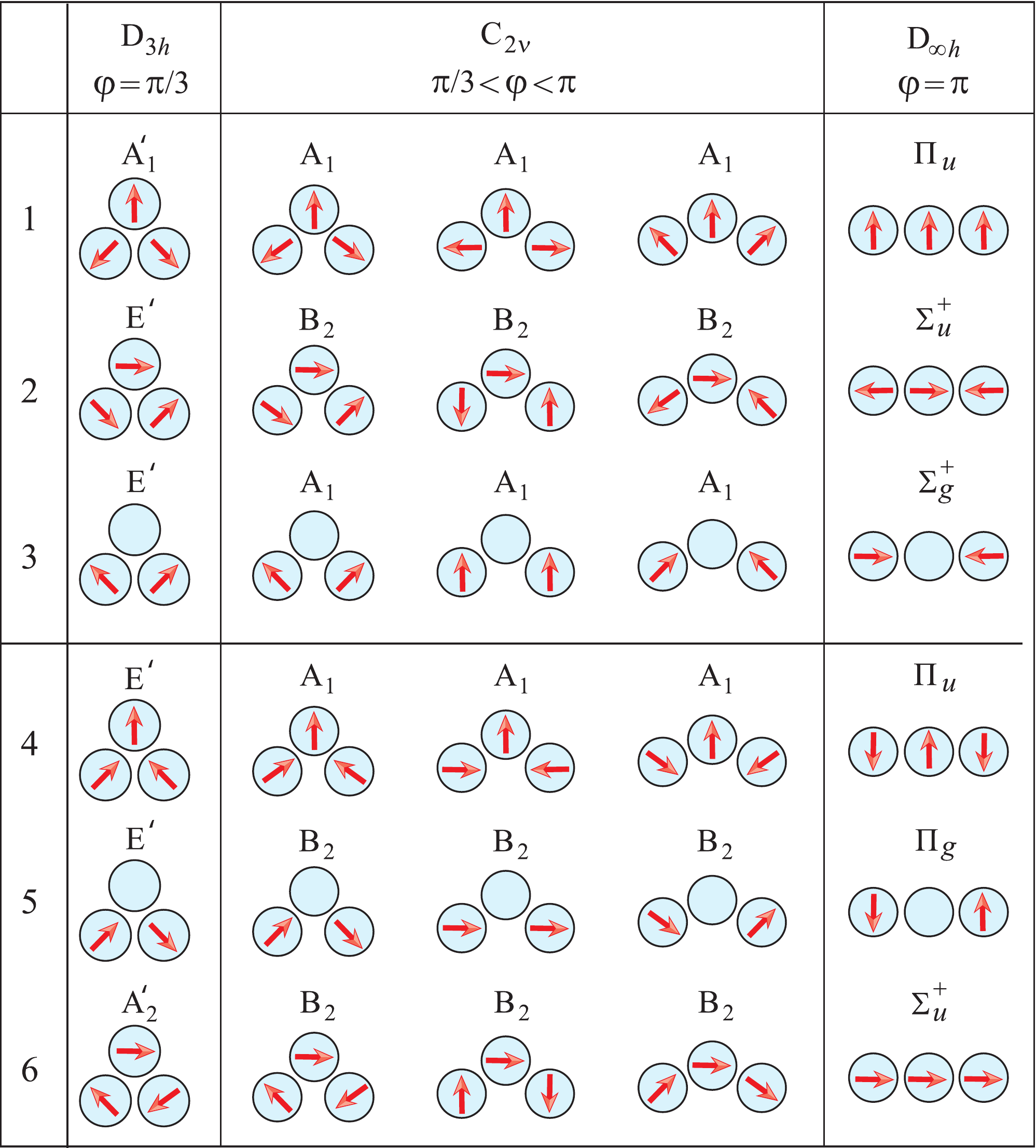}
\caption*{Table S1: The shapes of possible dipolar plasmon modes of trimers at varying vertex angle $\pi/3\le \varphi \le \pi$ in accordance with plasmon hybridization theory \cite{Chuntonov_NanoLett_2011}. Upper and lower parts correspond to antibonding and bonding modes, respectively.}
\label{fig:s1}
\end{figure}

\begin{figure}[ht!]
\centering\includegraphics[width=10cm]{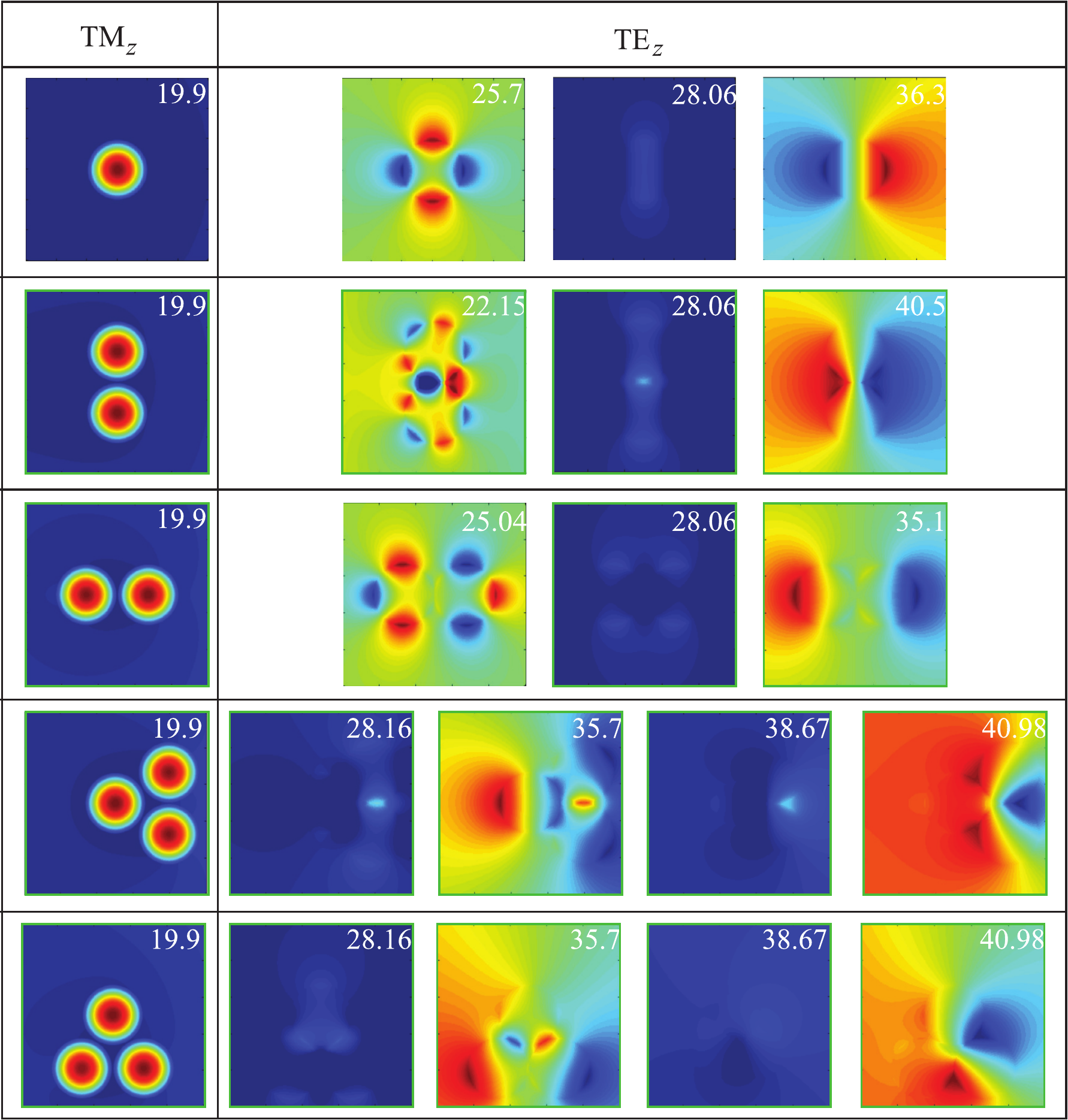}
\caption*{Figure S2: Near-field patterns of a single graphene-coated nanowire, dimer and symmetric trimer. Incidence of both TM$_z$-polarized (left column) and TE$_z$-polarized (right column) plane wave is present. All parameters of the problem are the same as in Figure 2 of the manuscript. The corresponding wavelengths in $\mu$m are given in the upper right corners of each field pattern}
\label{fig:s2}
\end{figure}

\begin{figure}[ht!]
\centering\includegraphics[width=10cm]{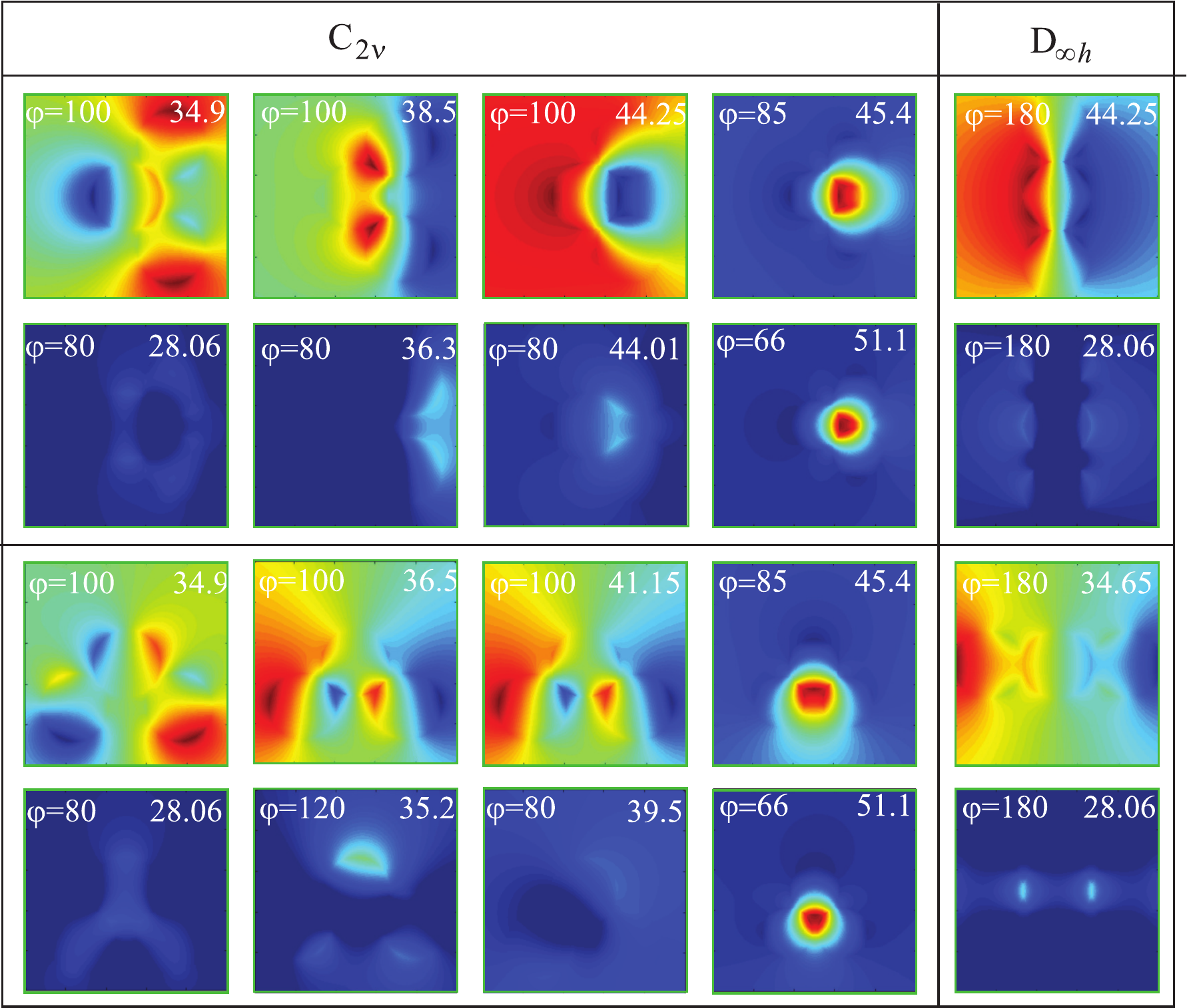}
\caption*{Figure S3: Near-field patterns of the asymmetric graphene-coated nanowire trimer. TE$_z$-polarized incident plane wave is considered. Patterns correspond to several selected points on the $\varphi - \lambda$ plane shown in Figure 4 of the manuscript. The corresponding wavelengths in $\mu$m are given in the upper right corners of each field pattern.}
\label{fig:s3}
\end{figure}
\end{document}